\documentclass[a4paper,fleqn]{cas-dc}

\usepackage[numbers,sort&compress]{natbib}
\usepackage{subfigure}
\makeatletter
\renewcommand{\fnum@figure}{Fig.\thefigure}
\makeatother
\def\tsc#1{\csdef{#1}{\textsc{\lowercase{#1}}\xspace}}
\tsc{WGM}
\tsc{QE}

\begin{document}
\let\WriteBookmarks\relax
\def\floatpagepagefraction{1}
\def\textpagefraction{.001}


\shortauthors{Prabhu et al.}  

\title [mode = title]{AI-based Carcinoma Detection and Classification Using Histopathological Images: A Systematic Review}  

\author[aut1]{Swathi Prabhu}
\fnmark[1]
\ead{prabhuswathi2@gmail.com}
\affiliation[aut1]{organization={Manipal School of Information Sciences}, addressline={Manipal Academy of Higher Education}, city={Manipal}, postcode={576104}, state={Karntaka}, country={India}}

\author[aut1]{Keerthana Prasad}
\cormark[1]
\cortext[cor1]{Corresponding author}
\fnmark[1]
\ead{keerthana.prasad@manipal.edu}

\author[aut2]{Antonio Robels-Kelly}
\affiliation[aut2]{organization={School of Information Technology}, addressline={Faculty of Science Engineering and Built Environment, Deakin University}, city={Geelong}, postcode={VIC 3216}, state={Victoria}, country={Australia}}
\ead{antonio.robles-kelly@deakin.edu.au}
\fnmark[2]            
\author[aut2]{Xuequan Lu}
\ead{xuequan.lu@deakin.edu.au}
\fnmark[2]



\begin{abstract}
Histopathological image analysis is the gold standard to diagnose cancer. Carcinoma is a subtype of cancer that constitutes more than 80\% of all cancer cases. Squamous cell carcinoma and adenocarcinoma are two major subtypes of carcinoma, diagnosed by microscopic study of biopsy slides. However, manual microscopic evaluation is a subjective and time-consuming process. Many researchers have reported methods to automate carcinoma detection and classification. The increasing use of artificial intelligence (AI) in the automation of carcinoma diagnosis also reveals a significant rise in the use of deep network models. In this systematic literature review, we present a comprehensive review of the state-of-the-art approaches reported in carcinoma diagnosis using histopathological images. Studies are selected from well-known databases with strict inclusion/exclusion criteria. We have categorized the articles and recapitulated their methods based on specific organs of carcinoma origin. Further, we have summarized pertinent literature on AI methods, highlighted critical challenges and limitations, and provided insights on future research direction in automated carcinoma diagnosis. Out of 101 articles selected, most of the studies experimented on private datasets with varied image sizes, obtaining accuracy between 63\% and 100\%. Overall, this review highlights the need for a generalized AI-based carcinoma diagnostic system. Additionally, it is desirable to have accountable approaches to extract microscopic features from images of multiple magnifications that should mimic pathologists$'$ evaluations.
\end{abstract}


\maketitle
\section{Introduction}\label{}
\label{int}

Cancer is a leading cause of death worldwide \cite{1}. The International Agency for Research on Cancer reported \cite{1,2} nearly 18.1 million new cases and 9.6 million cancer deaths worldwide in 2018. It has risen to 19.3 million cases and 10 million cancer deaths in 2020. Carcinoma accounts for 80 to 90 percent of all cancer cases that emerge in epithelial cells. The two major subtypes of carcinomas are adenocarcinoma (ADC) and squamous cell carcinoma (SCC). ADC occurs in mucus-producing glandular cells including lung, pancreas, esophagus, colon, breast, prostate, cervix, and vagina. SCC develops in multiple regions such as head and neck, lung, cervix, esophagus, prostate, vagina, and skin \cite{3}.

Early detection and treatment of carcinoma can reduce death rates and improve long-term survival rates. Conventional diagnosis of carcinoma involves clinical examination, laboratory tests, imaging tests (CT, MRI, PET, X-ray), and biopsy \cite{4,5}. Biopsy is the gold standard diagnostic technique used to remove tissue samples from the affected area. The scientific examination of tissue samples under a microscope after the staining process is called histopathology. Hematoxylin and eosin (H\&E) staining is a routine staining procedure used by pathologists. As a result of treating the cells with the hematoxylin stain, nuclei of the cells turn a purplish blue color, and cytoplasm and connective tissue turn pink due to the eosin stain \cite{5}. 

The purpose of histopathological image analysis is to examine tissue growth patterns and cell morphology. Pathologists scan tissue slides under a microscope at multiple magnifications. At low magnifications (4X, 10X), they observe the overall tissue sample and get a perspective about the malignant region, especially the architectural pattern of malignancy. Subsequently, they carefully study the slides at higher magnifications (40X, 60X, 100X), to evaluate the morphology at the cellular level. Generally, any malignant tumor has some common microscopic features, such as irregular shape and size of cells and nuclei, hyperchromatic (dark) nuclei, prominent nucleoli, increased nucleus to cytoplasm ratio (NC ratio), increased number of mitotic figures, and abnormal mitotic divisions. \cite{5}.

\begin{figure}[!ht]
    \centering
    \includegraphics[width=85mm,scale=1.2]{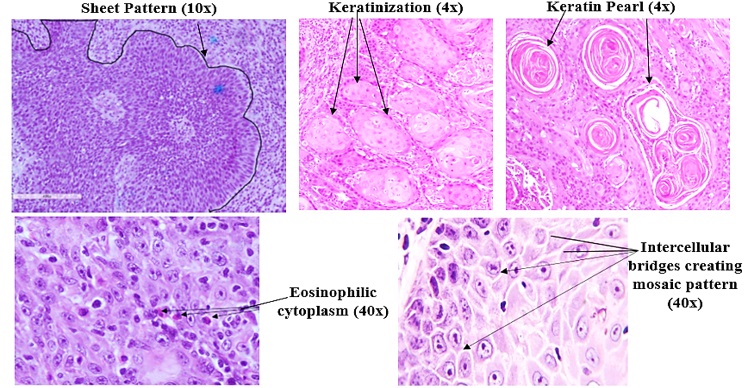}
    \caption{Histopathological features of SCC}
    \label{figure0}
\end{figure}

\begin{figure}
    \centering
    \includegraphics[width=80mm,scale=1.2]{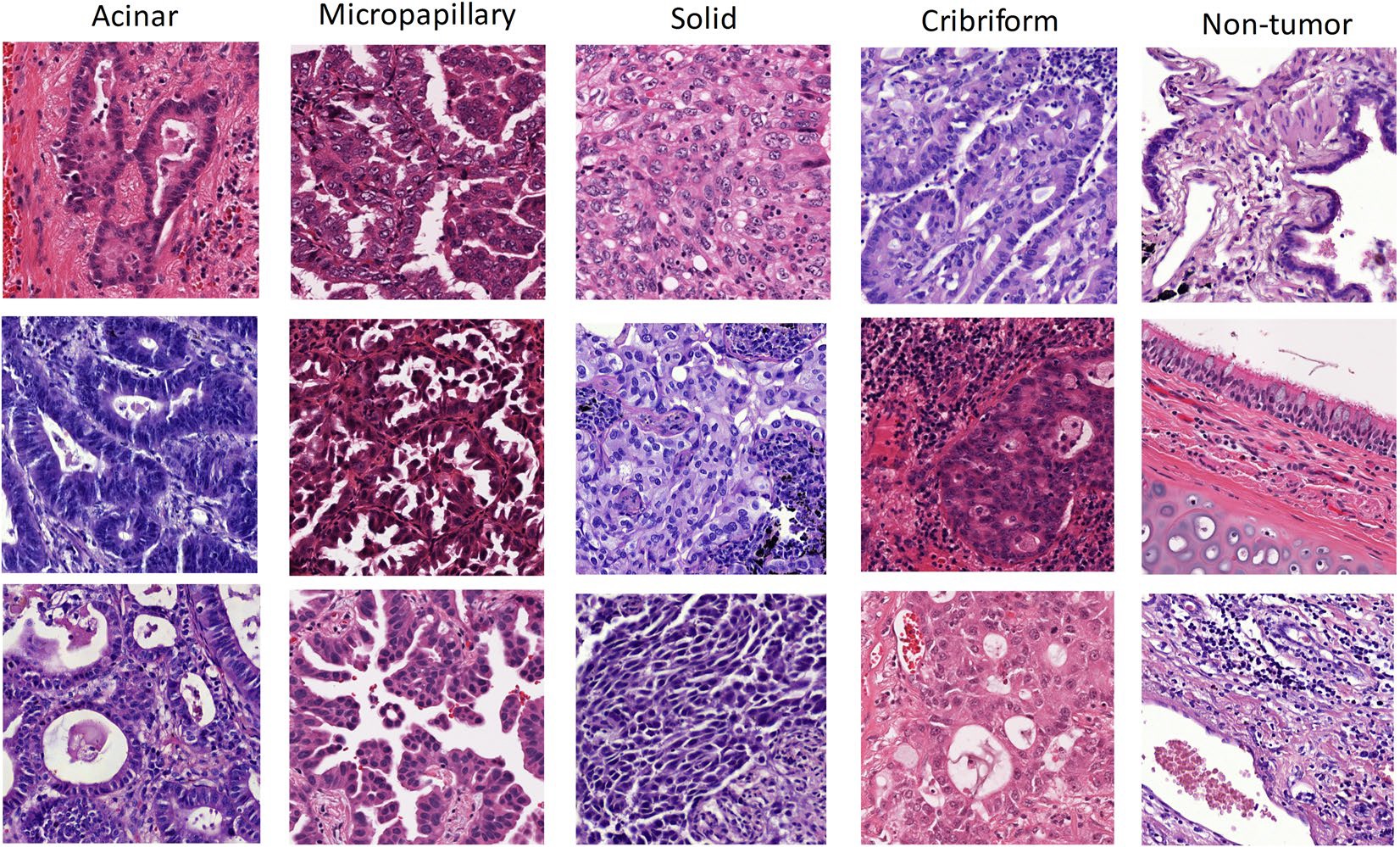}
    \caption{Different histopathological patterns of ADC [8]}
    \label{figure1}
\end{figure}

Histopathological characteristics of SCC include sheets or clusters of polyhedral cells, keratinization, and keratin pearl formation. Additionally, the presence of large cells, vesicular nuclei, and abundant eosinophilic cytoplasm are some of the other microscopic features \cite{6, 7}. Intercellular bridges are a characteristic of squamous epithelium and can be observed at the early stages of SCC \cite{9}. Fig.\ref{figure0} shows a few SCC microscopic features at multiple magnifications. ADC shows lepidic, acinar, papillary, micropapillary, solid, and cribriform histopathological growth patterns in some organs, as shown in Fig.\ref{figure1} \cite{10}. Histopathological variants of SCC and ADC can be differentiated using cell level information \cite{11}.

Manual histopathological examination is laborious, time-consuming, and subjective, with considerable inter and intra-observer variations among experts \cite{4,5}. These limitations call for automation of clinical decision-making. With the technological advancement over the past decades, many researchers have combined image processing, pattern recognition, machine learning (ML), and deep learning (DL) techniques to develop computer-aided diagnostic (CAD) systems for carcinoma detection and classification in order to achieve rapid and reproducible results \cite{12}.

ML is a core discipline of AI that learns discriminative patterns from existing data \cite{4, 12}. However, this requires human efforts and domain knowledge to extract appropriate features. It is also important to apply preprocessing algorithms to improve feature extraction, by suppressing unwanted distortions and enhancing or normalizing the histopathological images. DL is considered as a recent evolution of ML with complex functionality. In order to improve the performance of CAD systems, DL extracts the features directly from raw images instead of feature engineering. ML/DL algorithms improve the CAD system performance with increased training samples because that enables the computer to learn from experience \cite{4}. The classical ML/DL approaches include segmentation of region of interest (ROI), carcinoma detection, and carcinoma classification as different grades or types of carcinoma. These approaches either use microscopic images or whole slide images (WSIs) and extract local or global features from images for diagnosis. Thus, in cancer research, accurate detection and differentiation among carcinomas are crucial for appropriate treatment decisions.

Several studies have demonstrated the potential of ML- and DL-based methods for ADC and SCC detection and classification that complement the conventional light microscopy analysis by experienced pathologists. The main challenge in the histopathological research field is to build a novel, generalized, and fully automatic CAD system that provides an appropriate diagnosis robust enough to handle either microscopic images or WSIs, irrespective of imaging artifacts. In this article, we present a systematic review of the literature to evaluate the application and performance of ML and DL algorithms in detection and classification of malignant lesions of epithelial layer. We also highlight our observations on research gaps and make recommendations for future research. 

We organize our review article as follows.  Section \ref{sec2} provides methodology for the systematic literature review. Sections \ref{sec3} and \ref{sec4} present search results and article analysis, respectively, followed by summary of SCC and ADC in Section \ref{sec5} and \ref{sec6}, respectively. Section \ref{sec7} provides discussion and future direction, and Section \ref{sec8} presents the conclusion to the article.

\section{Methodology for the systematic literature review (SLR)}
\label{sec2}

A systematic approach is used to identify the literature in this study. The methodology includes defining a problem, considering articles, which satisfy inclusion/exclusion criteria, extracting relevant information from the article, and finally, assessing the extracted data \cite{13}. The main aim of this study is to explore the literature based on the terms that represent the primary concepts: ``squamous cell carcinoma,'' ``adenocarcinoma,'' ``automated detection and classification,'' and ``histopathology.'' SLR was carried out to track the technological development and to identify future scope in carcinoma detection and classification. We conducted the literature search between December 2019 and May 2021.

\subsection{Study selection}
The literature contribution was found by exploring iterative and systematic search on online databases, namely Scopus, Web of Science, IEEE Xplore digital library, ScienceDirect, PubMed, and ACM. The search strings considered were ((``Deep learning'' OR ``artificial intelligen*'' OR ``machine learning'' OR ``neural network'' OR ``image process*'') AND (``squamous cell carcinoma'') AND (``histopath*'' OR ``H\&E'')). The same search string was used to find ADC articles by replacing the phrase``squamous cell carcinoma'' with ``adenocarcinoma.'' This search string was jointly developed by the authorship team after examining the terminology used in different articles. The research questions of this study are listed below

The research questions of this study are listed below. 
\begin{enumerate}
\item What has been the research trend in automated detection and classification of carcinoma?
\item What type of preprocessing and segmentation methods were involved in current literature?
\item What methods were used for classification and detection of carcinoma of various organs? 
\item What kind of histopathological imaging and magnifications were reported in these studies?
\end{enumerate}

\underline{Inclusion criteria}
\begin{itemize}
\item  Studies using ML and DL for automated classification, detection and grading of ADC and SCC
\item Studies using H\&E stained microscopic images and WSIs
\item Studies published in first and second quartiles (Q1 and Q2)  
\end{itemize}

\underline{Exclusion criteria}
\begin{itemize}
\item Studies using ML and DL model to predict the carcinoma prognosis, recurrence, survival rate, metastasis, and treatment efficacy
\item Studies using ML and DL models for detection of other carcinomas
\item Studies that include search string in indexed keywords of the articles
\item Conference articles, letter to editors
\item Studies reported in languages other than English 
\end{itemize}

\underline{Information obtained from articles}
We extracted the information from selected articles, analyzed, and tabulated it in the Microsoft Excel format. The following details were noted down:
\begin{itemize}
\item Publication details (article title, authors name, year, country, problem statement)
\item Dataset details and image magnification used for study
\item Methods used in study (ML algorithms, DL models)
\item Outcome in terms of accuracy, precision, recall, F1 score, area under curve (AUC)
\end{itemize}

\section{Search Results}
\label{sec3}
We limited the article search to the article title, abstract, and author\textquotesingle s keywords. A total count of 2,579 articles was obtained. Search discipline was not limited to any domain. Subsequently, 86 articles were selected in accordance with inclusion and exclusion criteria and after the removal of duplicated articles among the databases. The reference list of the selected articles was considered and screened. A total of 15 articles that were not being selected from the search string was also considered for the study. Finally, a comprehensive full-text examination was performed on 101 journal articles. Table \ref{table1} summarizes article search results.

\begin{table*}[hbt]
\setlength{\tabcolsep}{1.5pc}
\caption{Article search results}
\label{table1}
\begin{tabular*}{\textwidth}{@{}lccccccc}
\hline Repositories &
                 \multicolumn{2}{m{4cm}}{Total papers based on title, abstract and keywords} 
                 & \multicolumn{2}{m{4cm}}{After applying inclusion exclusion criteria and removing duplication}
                 & \multicolumn{2}{m{3.0cm}}{Articles selected from the reference list of selected articles}
                 \\
                 & \multicolumn{1}{c}{ADC} 
                 & \multicolumn{1}{c}{SCC} 
                 & \multicolumn{1}{c}{ADC} 
                 & \multicolumn{1}{c}{SCC}    
                  & \multicolumn{1}{c}{ADC} 
                 & \multicolumn{1}{c}{SCC}  \\
\hline
Scopus	& 563 &	310	&26	& 29 &  \multirow{6}{*}{7} &  \multirow{6}{*}{8} \\ 
PubMed	& 246 &	157 & 4	& 5 & & \\
ACM digital library & 	10 & 3 &	0	& 0 & & \\
IEEE Xplore &	22 &	10 &	1 &	1 & &\\ 
Web of Science &	127	& 85 &	5 & 4 & & \\ 
ScienceDirect & 648 &	398	 & 5 & 6 & & \\  
\textbf{Total }&\textbf{1616} &	\textbf{963} &	\textbf{41} &\textbf{45} &	\textbf{7} & \textbf{8}\\  
\multicolumn{7}{l}{\textbf{Final selected articles =101}} \\
\hline
\end{tabular*}
\end{table*}

\begin{figure*}
    \centering
    \includegraphics[width=110mm,height=5in]{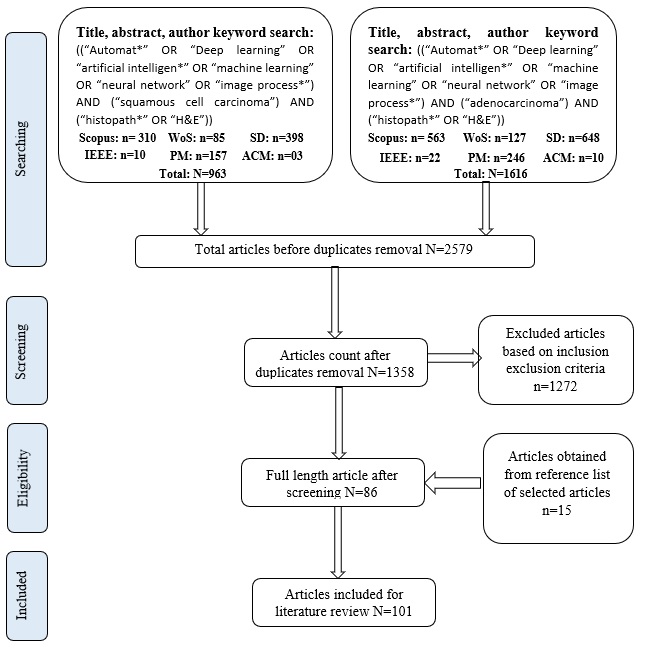}
    \caption{Study selection process}
    \label{figure2}
\end{figure*}

The schematic representation of the systematic review process is shown in Fig.\ref{figure2}. In the subsequent section, we discuss the analysis performed in this study.

\section{Analysis}
\label{sec4}
In sections \ref{sub1} and \ref{sub2}, we discuss the relevant articles to identify the current trends and technology of carcinoma detection and classification. The descriptive analysis and article analysis were performed to investigate the importance of the topic.

\subsection{Descriptive analysis}
\label{sub1}
Descriptive analysis was carried out to observe the statistics related to the articles obtained during SLR. We conducted the quantitative analysis based on the article information to assess the relevance of our domain of interest. The descriptive analysis of selected articles was performed based on the year of publication, leading journals, and geographical context. Sections \ref{year}, \ref{rank}, and \ref{country} provide a comprehensive overview of descriptive analysis.

\subsubsection{Year-based analysis}
\label{year}
\begin{figure}
    \centering
    \includegraphics[width=85mm,scale=1.7]{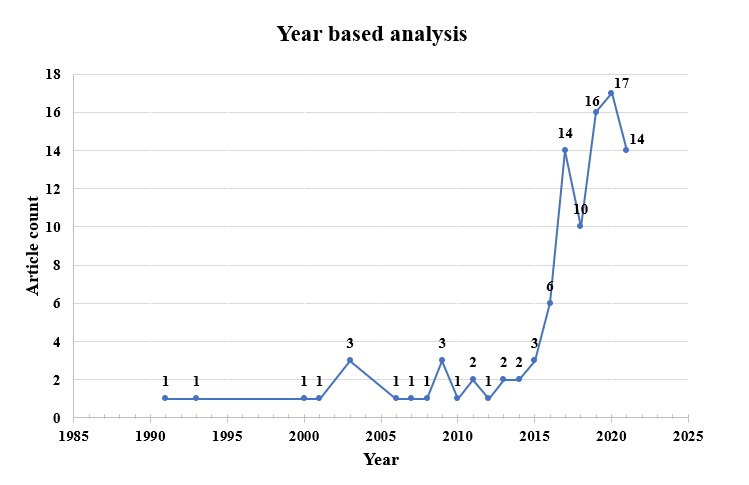}
    \caption{Article analysis based on year of publication}
    \label{figure3}
\end{figure}

In Fig.\ref{figure3}, it can be observed that carcinoma is one of the study areas since 1990. However, those studies were heavily dependent on the statistical analysis. Many of the 101 articles reported have been written in the last decade because of rapid evolution in ML and DL techniques and their applications in carcinoma diagnosis. Based on the trend, we anticipate that a greater number of studies could emerge within the research domains of automation systems for carcinoma detection and classification.

\subsubsection{SJR/JCR quartile-based analysis}
\label{rank}
\begin{figure}
    \centering
    \includegraphics[width=85mm,scale=3.9]{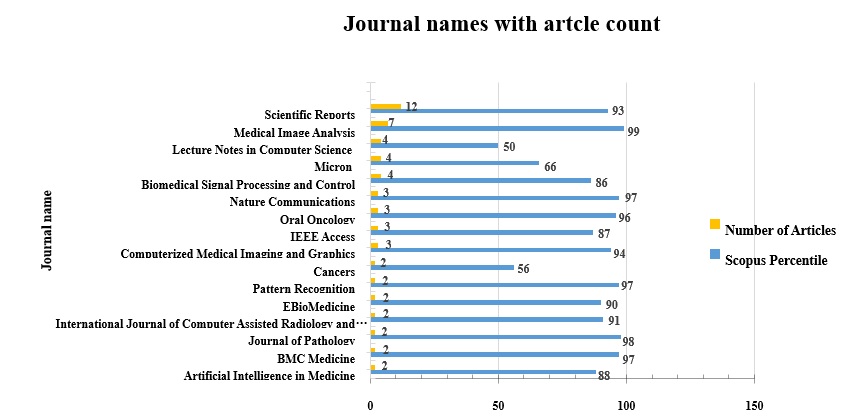}
    \caption{Analysis based on article count with journal names}
    \label{figure4}
\end{figure}

We divided the journals into each subject category based on quartiles of Journal Citation Reports (JCR)/SCIMAGO Journal and Country Rank (SJR). Out of 60 journals, the top 16 journal names that published greater than or equal to two articles and the article counts of the corresponding journals are depicted in Fig.\ref{figure4}. We carefully chose articles from Q1 and Q2 because the most prestigious journals within a subject area occupy Q1 (top 25\% of journals in the list) and Q2 (top 25\% to 50\% group). Among the selected articles, 76.6\% (46 out of 60 journals) of the articles were published in Q1 journals and 23.4\% (14 out of 60 journals) of the articles were published in Q2 journals. Additionally, among 60 journals, 30\% (18 journal) of journals were of open access type. The year of quartile rank list consideration is 2020.

\subsubsection{Geographical context-based analysis}
\label{country}
\begin{figure}
    \centering
    \includegraphics[width=85mm,scale=7.0]{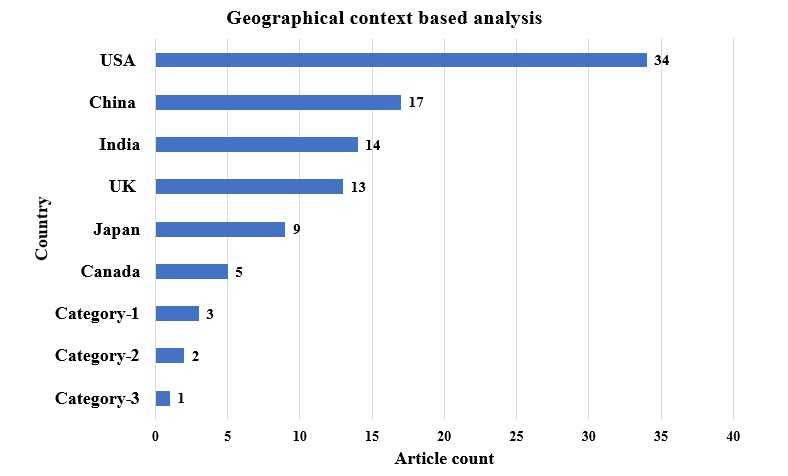}
    \caption{Country-based analysis}
    \label{figure5}
\end{figure}

 Fig.\ref{figure5} depicts geographic context-based analysis of carcinoma detection and classification, where it shows article count of each country. We grouped some couturiers into category-1, category-2, and category-3 based on their article count as 3, 2, and 1, respectively. Category-1 in Fig.\ref{figure5} includes Netherlands, Turkey, Switzerland, Iran, France, Brazil, and Singapore. Category-2 comprises Sweden, Finland, Italy, South Africa, Austria, Germany, Israel, and Pakistan. Category-3 includes Malaysia, Libya, Republic of Korea, Saudi Arabia, Slovenia, Spain, Sri Lanka, Taiwan, Bangladesh, and Croatia. The majority of research was reported by developed and developing economies, with the top four countries on the list being the United States (34 papers), China (17 papers), India (14 papers), and the United Kingdom (13 papers). The articles also show collaborative work with multiple countries, which indicates the importance of research on carcinoma in a global scenario. Fig.\ref{figure6} shows the count of articles that has been published by various universities of the top four countries. We observed the authors$'$ collaborative work with different universities within these countries and among other countries too. It clearly shows the widespread interest among people from various universities.

\begin{figure*}
    \centering
    \mbox{\subfigure{\includegraphics[width=8.5cm, height=5.5cm]{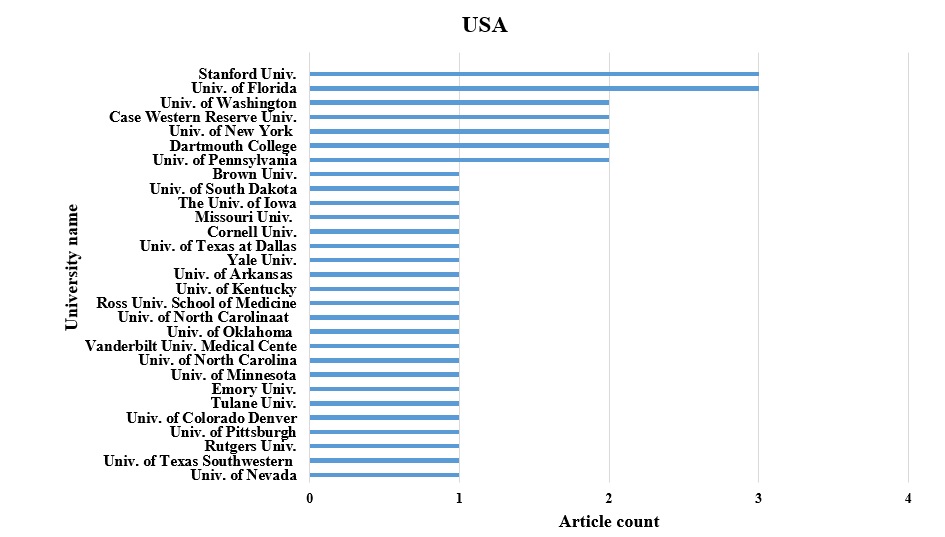}}\quad
        \subfigure{\includegraphics[width=8.0cm,height=4.5cm]{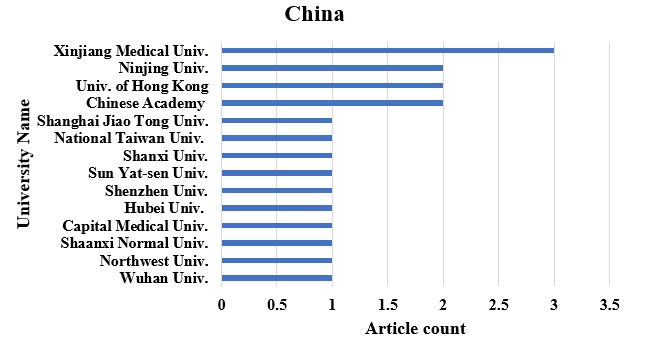}}}
\end{figure*}
\begin{figure*}
    \centering
    \mbox{\subfigure{\includegraphics[width=8.5cm, height=4.8cm]{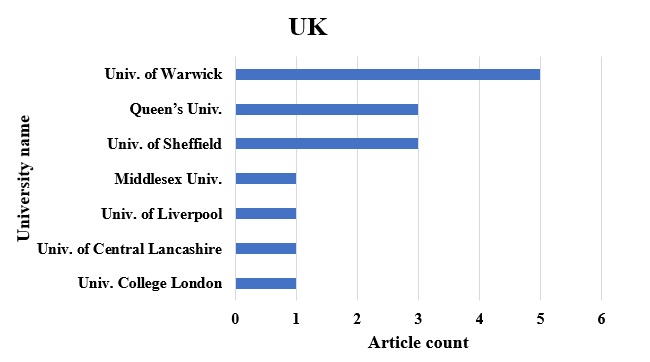}}\quad
        \subfigure{\includegraphics[width=8.0cm,height=4.5cm]{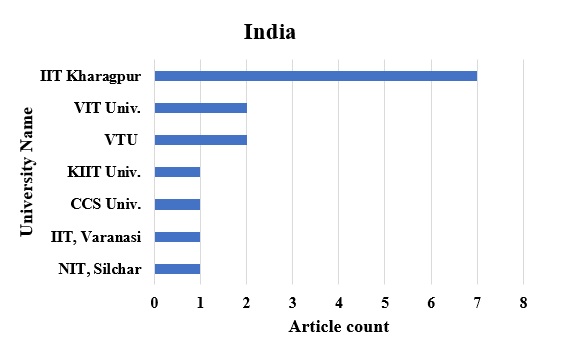}}}
    \caption{University-based analysis in top four countries}
    \label{figure6}
\end{figure*}

\subsection{Article analysis}
\label{sub2}
\begin{figure*}
    \centering
    \includegraphics[width=95mm,scale=1.0]{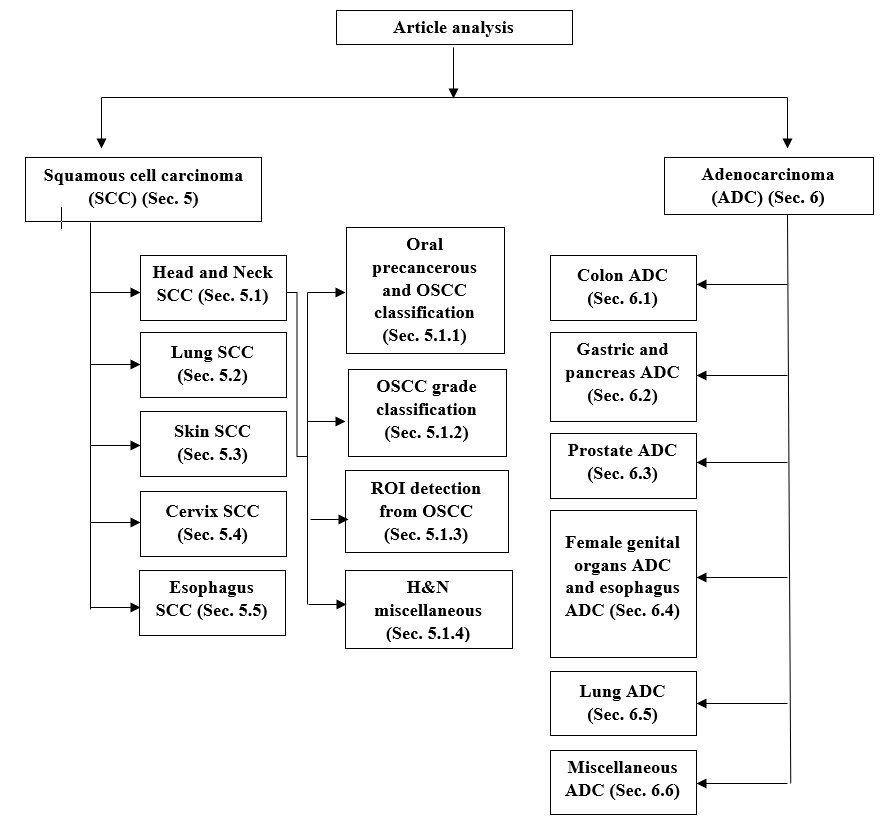}
    \caption{Section summary}
    \label{figure7}
\end{figure*}

Our main objective is to group the downloaded articles in such a way that each group provides essential information about carcinoma diagnosis. We consider various research articles to provide a comprehensive review of AI-based techniques for carcinoma diagnosis using histopathological images. Fig.\ref{figure7} shows categorization of articles based on the primary organ of carcinoma, and its detailed discussion is provided in Sections \ref{sec5} and Section \ref{sec6}.

\section{Squamous Cell Carcinoma}
\label{sec5}
In this section, we describe different approaches used to detect and classify SCC of various organs. Sections \ref{scc1}, \ref{scc2}, \ref{scc3}, \ref{scc4}, and \ref{scc5} provide a detailed discussion on the existing literature for SCC diagnosis of head and neck region, lung, skin, cervix, and esophagus, respectively.

\subsection{Head and Neck SCC}
\label{scc1}
SCC of head and neck (H\&N) i is the sixth leading cancer incidence worldwide, with approximately 600,000 new cases annually \cite{14}. The prevalence rate is increasing in Southeast Asian countries due to alcohol consumption, smoked and smokeless tobacco, diet, viruses, immunosuppression, and chronic infections. H\&N SCC develops in mucosal linings of upper aerodigestive tract, tongue, oral cavity, pharynx, larynx, trachea, nasal cavity, and paranasal sinuses \cite{4}. Oral SCC (OSCC) was observed in 95\% cases of oral cancer and is the most common among H\&N region \cite{15}. The higher incidence of oral cancer is due to the late diagnosis of potential precancerous lesions. Section \ref{sccsub1} discusses SCC and precancerous lesion classification. Section \ref{sccsub2} describes different grades of OSCC. Section \ref{sccsub3} reports ROI detection from the tissue sample that helps SCC diagnosis, and Section \ref{sccsub4} provides the literature on SCC of H\&N regions other than oral cavity.

\subsubsection{Oral precancerous and cancerous lesion classification}
\label{sccsub1}
Diagnosis of precancerous lesions of oral cavity, namely oral submucous fibrosis, leukoplakia, and erythroplakia, at the earliest is important because they could progress into carcinoma in situ and/or SCC \cite{16}. Mahmood et al. \cite{4}, Nag et al. \cite{17}, and Alabi et al. \cite{101} provided a comprehensive review on precancerous and cancerous lesions of H\&N region. Precancerous changes manifest significant cytological and tissue architectural changes such as pleomorphism, increased NC ratio, loss of polarity of basal layer, increased mitotic count, keratinization in single cells, keratin pearls, and irregular epithelial stratification \cite{17}. Nag et al. \cite{17} highlighted the importance of tissue architecture and analysis of cellular changes for early detection that helps improve patient stratification. 

In \cite{16}, Liu et al. proposed oral cancer risk index that helps prediction of OSCC risk in patients with oral leukoplakia. The specificity and sensitivity of 99\% and 98\% were achieved using support vector machine (SVM) classifier. Krishnan et al. \cite{18} segmented the epithelium layer based on texture and color gradient. They obtained 98\% segmentation accuracy. Further, in \cite{22, 19, 21}, the authors classified the oral tissue images as normal, oral precancerous lesion, and OSCC. Krishnan et al. \cite{22} differentiated oral tissue images based on texture features of collagen fibers using SVM and obtained 91.64\% accuracy. In \cite{21}, Krishnan et al. calculated epithelium thickness and tissue architecture using fractal dimension and performed statistical analysis to classify the images. Krishnan et al. \cite{19} extracted HOS-, LBP-, and LTE-based texture features from histopathological images. Out of the five classifiers, fuzzy classifier achieved 95.7\% accuracy. These studies were carried out with the images captured at 40X \cite{22} and 10X \cite{18, 19, 21} magnifications. Many researchers reported the feature extraction from epithelial tissue \cite{22, 19, 18, 21}. However, it is also important to measure the epithelial-connective tissue interface (ECTI) for proper diagnosis. Eid et al. \cite{20} used fractal geometry quantifiers to quantify ECTI irregularity, using 377 histopathological images collected at 40X magnification. They used linear discriminant analysis (LDA) for classification and reported a sensitivity and specificity of 63\% and 67\%, respectively. Along with tissue architecture analysis, cell nuclei were considered to assess tumor development by Landini et al. \cite{23} based on graph theory. The discrimination rates of 67\%, 100\%, and 80\% were obtained for normal, dysplastic, and OSCC cells, respectively.

\subsubsection{Oral cancer grade classification}
\label{sccsub2}
Carcinomas can be graded as well differentiated (WD), moderately differentiated (MD), and poorly differentiated (PD). Almangush et al. \cite{24} discussed different WHO grading and staging systems, and highlighted the importance of AI to improve staging, grading, and treatment planning of OSCC. Rahman et al. \cite{25} and Nawandhar et al.\cite{15} proposed a ML-based classifier to classify normal cells and oral SCC. Further, they graded OSCC into three grades based on texture features, gradient features, geometrical features, and Tamura features. In \cite{15}, Nawandhar et al. classified 676 OSCC microscopic images using maximum voting method and obtained 95.56\% accuracy. A total of 305 features were extracted after applying contrast limited adaptive histogram equalization (CLAHE) technique. Rahman et al. \cite{25} considered texture-based features for SVM classifier to distinguish normal and cancerous cells and achieved 100\% accuracy. Although both groups of researchers collected the images at 40X magnification, the experiments were conducted on different image datasets. Das et al. \cite{26} extracted small patches from 156 WSIs, digitized at 40X magnification instead of digital microscopic images for OSCC grade classification. They performed a comparative study among the transfer learning approach using Alexnet, VGGNet, and ResNet models with that of the training from scratch. The results showed that convolutional neural network (CNN) designed from scratch performed better than transfer learning models, with an accuracy of 97.51\%. In \cite{98}, Musulin et al. exploited XceptionNet with stationary wavelet transformation technique to grade 322 images captured at 10X magnification. They obtained a classification AUC of 0.96.

\subsubsection{ROI detection from OSCC}
\label{sccsub3}
OSCC WSIs were classified as tumor, lymphocytes, and stroma using ResNet, DenseNet, InceptionNet, XceptionNet, and MobileNet, out of which MobileNet showed superior performance, with an accuracy of 96.31\% \cite{27}. A few histopathological features such as nuclei detection and delineation \cite{28, 29, 97}, mitotic cell detection and count \cite{30, 31}, tumor-associated tissue eosinophilia \cite{32}, microvessel architecture \cite{33, 34}, and keratin pearl count \cite{34} were used to distinguish OSCC from various other types of cancer. GPU-based squamous cell segmentation was performed by Nawandhar et al. \cite{97}, using k-means clustering algorithm. They extracted morphological and Law\textquotesingle s texture features from images of SCC cases. Lu et al. \cite{29} reported nuclei segmentation using watershed method and extracted shape features and texture features from segmented nuclei. The extracted features were statistically analyzed using Wilcoxon rank sum test to identify the top five histomorphometry features for further analysis. In \cite{28}, Das et al. proposed a two-stage model for automatic detection and segmentation of nuclei from OSCC. They implemented CNN model for nuclei detection and achieved precision of 82.03\%. In the second stage, the authors used Chan-Vese model for segmentation of nuclei after applying contour enhancement using non-subsampled contourlet transform. Further, they also implemented an automatic technique to classify mitotic and non-mitotic cells \cite{30}. Stain normalization was performed using singular value decomposition (SVD) method and mitotic, and non-mitotic cells were extracted using Otsu\textquotesingle s thresholding. The statistical features, namely entropy measures, fractal dimension, and Hu\textquotesingle s moment-based descriptors, were used for random forest (RF) tree ensemble learning algorithm to classify mitotic and non-mitotic cells \cite{30}. Tandon et al. \cite{31} presented a study on comparison of mitotic figure count in normal oral sample, oral dysplasia, and oral SCC tissue sample stained with crystal violet and H\&E staining technique. SCC group had a significantly higher proportion of abnormal mitotic figures compared with normal and dysplasia samples irrespective of the stain used. Lorena et al. \cite{32} used immunohistochemistry (IHC) and H\&E stained images to count eosinophils in oral SCC images. Statistical tests supported the evidence that special staining techniques such as IHC were not required to identify eosinophils in oral SCC. 

Microvessels in tumor tissue varied in terms of size, shape, and texture \cite{33, 34}. Fraz et al. \cite{33} segmented microvessels from WSIs of oral SCC samples using deep network and achieved an accuracy of 96\%. Tan et al. \cite{34} performed morphometric analysis of microvessel density (MVD) and keratin pearls of three different grades of OSCC images, captured at 20X and 10X magnifications. Morphometric analysis showed that MVD was significantly lower in WDSCC compared with MDSCC, and the keratin pearl count was significantly higher in WDSCC compared with MDSCC and PDSCC. Recently, deep segmentation models such as SegNet, UNet, and DeepLabv3+ were used to segment the tumor region from histopathological images \cite{98, 102}.

\subsubsection{H\&N miscellaneous}
\label{sccsub4}
In the 1990s, a few researchers \cite{35, 36} differentiated benign and malignant squamous tumors of H\&N region based on nuclei information. Cooper et al. \cite{35} differentiated larynx SCC samples as verrucous SCC and benign squamous papilloma based on morphometric analysis. Jin et al. \cite{36} reported that the area and the diameter of nuclei were increased in invasive SCC in comparison with normal mucosa and squamous papilloma.  Mete et al. \cite{37} and Halicek et al. \cite{14} proposed automation systems to detect H\&N SCC using WSIs and achieved accuracies of 96\% and 85\%, respectively. DBSCAN clustering algorithm detected malignant areas based on nuclei color and shape features in HSV color space. They classified the WSI patches as normal or malignant using SVM \cite{37}. Halicek et al. \cite{14} also used HSV color space for InceptionNet  architecture in SCC diagnosis.

\subsection{Lung SCC}
\label{scc2}
Lung cancer accounts for more than 1.4 million deaths worldwide annually \cite{38}. Classification of lung cancer subtypes as ADC, SCC, large cell carcinoma, and small cell carcinoma was based on morphological features of tumor tissue, particular tumor cells, and surrounding tissue environment \cite{11, 39}. Several researchers \cite{38, 39, 40, 41, 93, 96} have reported the use of traditional ML approaches such as SVM \cite{38, 96}, RF \cite{93}, AdaBoost, cw-Boost \cite{40} fast adaptive neural classifier \cite{41}, and statistical model \cite{39} to distinguish ADC and SCC. The classification was performed based on various features extracted either from cells, nuclei, or images. The features such as texture features, morphological features, Hu\textquotesingle s moments, GLGCM, wavelet transform, GLCM, LBP, GLDS, Markov random field, VGGNet features, HOG, GIST, LeNet features, and densitometric features were reported in the literature \cite{11, 38, 39, 96, 40, 41, 43}. Li et al. \cite{96} applied Lasso and Relief feature selection methods and LDA feature reduction technique on the set of features extracted from tissue samples. In \cite{93}, Nishio et al. classified the lung cancer subtypes using the homology-based feature extraction method. They obtained accuracy of 78.33\% with RF classifier, which was better than the accuracy of 70.83\% using texture features with RF classifier. A few researchers \cite{11, 42, 43} reported the implementation of hash-based image retrieval techniques to distinguish lung cancer subtypes. Hashing-based image retrieval with majority voting method achieved 87.3\% accuracy \cite{11, 42}, whereas 97.98\% classification accuracy was obtained on supervised graph hashing technique \cite{43}. Segmentation of nuclei using Otsu\textquotesingle s thresholding method was reported \cite{38, 41}. Some research groups proposed feature enhancement methods to improve classification accuracy \cite{40, 41, 96}.

Further, deep  CNN pretrained models such as InceptionNet, AlexNet, VGGNet, and ResNet were exclusively used in ADC and SCC classification \cite{44, 45, 46, 92, 94, 95}. Coudray et al. \cite{46} considered InceptionNet and obtained AUC of 0.97 for the cancer genome atlas (TCGA) image classification. However, the authors tested the model on independent datasets collected at 5X and 20X magnifications. The results showed that AUC around 0.834 to 0.861 was obtained using images of 20X magnification. An AUC around 0.871 to 0.928 was achieved on images of 5X magnification. Noorbakhsh et al. \cite{94} also reported the use of transfer learning with InceptionNet network for TCGA tiles classification. They obtained AUC around 0.87 for ADC and SCC samples of lung, esophagus, and cervix region. Yu et al. \cite{45} used InceptionNet, VGGNet, and AlexNet models for the classification and obtained AUC of 0.93 for VGGNet and InceptionNet. However, AlexNet achieved a smaller AUC of 0.75. Hence, the authors concluded that tumor type and subtype classification might require deep layer networks. In these studies, WSI was broken into varied non-overlapping patches \cite{45, 46}. In \cite{44}, Khosravi et al. considered random patches from TCGA WSI dataset and fine-tuned the parameters of InceptionNet model. The classification accuracy of 92\% was achieved. Kanavati et al. \cite{92} and Yang et al. \cite{95} introduced EfficientNet for lung cancer subtype classification and AUC above 0.95 was obtained. Both research groups trained the model using private and public WSIs, digitized at 10X \cite{92} and 20X \cite{95} magnifications. Before training, Otsu\textquotesingle s thresholding was used to remove the white background region. In a recent study, Chen et al. \cite{100} reported a new method for lung cancer subtype classification using annotation-free WSIs with a slide-level label. They exploited ResNet model and obtained ROC values of 0.96 and 0.94 for ADC and SCC classification, respectively.

\subsection{Skin SCC}
\label{scc3}
Skin SCC is the second most common skin cancer; it accounts for around 20\% of non-melanoma skin malignancies \cite{47}. Noroozi et al. \cite{47} proposed an automatic method using histopathological images to differentiate SCC in situ from actinic keratosis. A total of 30 images were collected at 60X magnification. They extracted intensity profiles of epidermis region for Bayesian classifier and obtained 92.1\% accuracy. The same group of researchers \cite{49} differentiated basal cell carcinoma from SCC using 33 images captured at 40X magnification in another study. They considered Z-transform features with SVM classifier and achieved an accuracy of 84\%.

\subsection{Cervix SCC}
\label{scc4}
Cervical cancer is the most common cause of death in women, with 80\% of cases observed in developing countries \cite{50, 51, 52}. Cervical intraepithelial neoplasia (CIN) is a pre-malignant lesion. If CIN is left untreated, it may progress to SCC of cervix \cite{52}. Several researchers \cite{50, 51, 52, 53, 54} classified CIN into CIN-I, CIN-II, and CIN-III grades, based on the severity of abnormal cells. Wang et al. \cite{52} reported a block-based coarse segmentation of squamous epithelium from WSIs digitized at low magnification. Further, SVM was used for classification based on texture features obtained from the squamous epithelium. Keenan et al. \cite{51} presented a method to construct Delaunay triangulation mesh of nuclei to perform the nuclei analysis.  Discriminant analysis showed that the classification accuracy of CIN-I, CIN-II, and CIN-III were 76.5\%, 62.3\%, and 98.7\%, respectively. Zhang et al. \cite{53} proposed a shape-based graph search method for nuclei segmentation from images. Almubarak et al. \cite{54} classified the segments of images using CNN and constructed a new feature vector from probability of all segments to classify the image. The authors obtained 77.2\% accuracy using logistic regression (LR) and RF tree classifiers. Zhang et al. \cite{99} presented stacked ensemble framework to classify tissue samples into low and high SCC grades. They obtained 90\% accuracy for ensemble learning and 89.17\% after applying  synthetic minority oversampling data augmentation technique. Further, Wu et al. \cite{50} proposed CNN model to classify cervical SCC into three different subtypes as keratinizing, nonkeratinizing, and basaloid SCC. They reported the use of 3,012 patches obtained from 502 images and obtained 108,432 patches after data augmentation. The model achieved classification accuracy of 93.33\% and 89.48\% for the augmented image group and original image group, respectively.

\subsection{Esophagus SCC}
\label{scc5}
The eighth most common malignancy in the world is esophagus carcinoma (EC). Esophagus SCC (ESCC) and esophagus adenocarcinoma (EAC) are two major variants of EC. In western countries, EAC is the most common variant, and ESCC is dominant in developing countries \cite{55}. Das et al. \cite{26} proposed CNN-based SCC grade classification of esophagus and oral tissue images. Hosseini et al. \cite{55} differentiated precancer of ESCC, called squamous cell dysplasia, from normal tissue. They considered the images of two different magnifications and enhanced the G component of RGB images using CLAHE technique. Fractal-based analysis was performed to measure geometrical complexity of images. The k-nearest neighbour (KNN) classifier provided an accuracy of 97.78\% for classification.

\section{Adenocarcinoma}
\label{sec6}
In this section, we provide the details of state-of-the-art methods of ADC detection and classification using ML and DL approaches. We subdivided the section into six subsections based on the primary organ of origin, namely colon, gastric and pancreas, prostate, female genital organs and esophagus, lung, and miscellaneous ADC. The details of the methods used in the literature for ADC diagnosis are provided in Sections \ref{adc1}-\ref{adc6}.

\subsection{Colon ADC}
\label{adc1}
Colon and stomach cancer are the causes of leading deaths in developed countries, with colon cancer ranking second in women and third in men during 2018 \cite{62, 64}. Colon ADC is the most common cancer, and diagnosis is based on the architectural appearance of gland formation and its morphology \cite{56}. Segmentation helps to localize information such as texture and spatial arrangement of cells specific to the glandular areas. Many researchers used MICCAI Gland Segmentation (GlaS) Challenge in 2015, \cite{57} releasing images with annotations in their studies to demonstrate gland segmentation \cite{56, 58, 59}. CNNs manifest great potential in differentiating diagnostic features of H\&E stained images. Manivannan et al. \cite{56} combined handcrafted features with features computed by CNN for gland segmentation. They predicted local patch labels using SVM and obtained Dice score above 0.85. Kainz et al. \cite{58} also segmented and classified glands as benign and malignant. The authors separated the hematoxylin and the eosin components using the color deconvolution method and observed relevant information of tissue structure in red channel of CLAHE enhanced images. They proposed CNN models, namely Separate-net and Object-net for segmentation and classification, respectively. Binder et al. \cite{59} presented multiple organ glands and stroma segmentation using dense UNet from GlaS dataset. They obtained a Dice coefficient of 0.92 for colon gland segmentation and 0.93 for stromal segmentation. They also segmented breast glands and stroma with Dice coefficient of 0.78 and 0.87, respectively. The model was originally trained on colon dataset. Another group of researchers used private dataset to segment the glands \cite{60, 61, 62}. 

Qaiser et al. \cite{62} proposed fast tumor segmentation by RF regression model on different WSI datasets. They combined deep convolutional and persistent homology features extracted from hematoxylin, eosin, and background channels. A precision of 0.77 was obtained. However, all the above studies used pixel-based annotation to perform the segmentation. Tosun et al. \cite{60} and Gunduz-Demir et al. \cite{61} reported the object-based segmentation algorithm for ROI segmentation. The authors \cite{61} differentiated false glands and true glands from segmented glands using decision tree (DT) classifier, and accuracy of 82.57\% was achieved. Tosun et al. \cite{60} reported accuracy of 94.89\% for segmentation of normal and cancerous regions from 16 biopsy samples captured at 5X magnification. 

In \cite{63}, Onder et al. extracted texture features using local histograms and co-occurrence matrices, and labeled images as normal and colon ADC using quasi-supervised texture labeling methodology. They achieved false and true positive rates up to 19\% and 88\%, respectively. Iizuka et al. \cite{64} and Kosaraju et al. \cite{65} proposed CNN models to differentiate ADC and adenoma tumor patches of colon and gastric WSIs. Kosaraju et al. \cite{65} performed noise removal by Gaussian blur smoothing operation, followed by thresholding to remove the background. The images were captured at three different magnifications in which combination of 20X and 5X magnification levels obtained around 93\% accuracy. Additionally, results showed that the TCGA stomach and colon carcinoma exhibit similar pathological patterns \cite{65}. Yoshida et al. \cite{66} developed an e-Pathologist system using multi-instance learning, using neural network classifier. They analyzed the automated system on 1,077 and 251 tissue samples from the two hospitals, digitizing the images at 40X magnification. The undetected rate of carcinoma, undetected rate of adenoma, and over-detected rate were 0\%, 9.3\%, and 9.9\%, respectively for dataset 1 and 0\%, 36.1\%, and 27.1\%, respectively, for dataset 2 \cite{66}. Zhou et al. \cite{114} performed classification and localization of colon ADC using weakly supervised deep CNN methods. Gaussian blur operation and Otsu\textquotesingle s thresholding were used in this study. Further, CNN features from image level and heatmap features from cell level were extracted. They achieved an accuracy of 94.6\% and 92\%, respectively, for TCGA and private dataset using RF classifier. Ghosh et al. \cite{113} implemented ensemble CNN model to detect colon ADC tumor region in histopathological images. They considered two datasets with images of 20X magnification. Instead of single-center dataset, the model provided better accuracy of 99.13\% when datasets were combined. Liang et al. \cite{115} presented a method to identify histopathological images of ADC using multi-scale feature fusion CNN-based shearlet transform technique. They achieved detection accuracy of 96\% for 8,000 image patches in the training dataset.

\subsection{Gastric and pancreas ADC}
\label{adc2}
Gastric ADC is also known as stomach ADC. It represents 90\% to 95\% of stomach cancer cases in developed countries \cite{64}. Yoshida et al. \cite{67} reported a CNN model to detect stomach ADC using WSIs. The images of 3,062 gastric biopsy specimens were captured at high and low magnifications. Yoshida et al. \cite{66, 67} compared the e-Pathologist software to detect gastric and colon ADC with expert classification results. Although the overall concordance rate of e-Pathologist was as low as 55.6\% for the three-tier classification, it identified 90.6\% of negative specimens. Sharma et al. \cite{68} presented a comparative study among AlexNet transfer learning model, customized CNN model, and RF classifier with handcrafted features for gastric carcinoma classification. They obtained better results for RF model in comparison with CNN model. In \cite{109}, Yu et al. performed multi-centric study for gastric ADC and normal classification. WSIs were collected from four different private hospitals, and images were digitized at 20X magnification. They performed color deconvolution operation and Otsu\textquotesingle s thresholding to remove the background, followed by neural network for classification. Langer et al. \cite{69} detected pancreatic ADC from 103 images captured at 45X magnification collected from mice. They applied watershed segmentation for nuclei extraction. Stromal segmentation was performed using color clustering techniques. RF and DT classifiers were used for the detection in which RF results were more promising than DT. Recently, Naito et al. \cite{108} proposed EfficientNet deep network model for pancreatic ADC and non-ADC classification. They performed transfer learning and obtained 0.984 AUC.

\subsection{Prostate ADC}
\label{adc3}
Prostate ADC is one of the most frequently diagnosed cancers in men, with a high rate of death worldwide \cite{70}. Detection of prostate ADC using multiscale analysis on WSIs was reported \cite{70, 71, 72}. Rashid et al. \cite{70} proposed two-stage detection technique to differentiate benign and malignant glands with an accuracy of 93\%. Glands were segmented using LDA and connected component techniques with the images captured at 5X magnification. RF classifier used morphological features from segmented regions to identify benign or probable malignant regions. In the second stage, images captured at 20X magnification were considered for nuclei segmentation. They used modified watershed algorithm for nuclei segmentation and extracted specific features for final glands classification. Doyle et al. \cite{71} automated the detection of prostate ADC using a hierarchical cascaded scheme, where they considered first order statistics, co-occurrence, and wavelet features. At each scale, pixel-wise Bayesian classifier was designed to obtain the likelihood scenes. Nir et al. \cite{72} developed an algorithm to detect cancer and to find the grading agreement among the pathologists. They explored the use of ML and DL techniques for classification of benign and malignant patches. The proposed method achieved 92\% accuracy in cancer detection and 79\% of accuracy in cancer grading. The Kappa value, which represents classifier\textquotesingle s overall grading agreement with the pathologists, was reported as 0.51. Leng et al. \cite{73} designed a regression model to estimate cancerous regions in WSI and achieved AUC of 0.951. They reported the use of colorimetric analysis of H\&E and IHC stained histopathological specimens.

Gleason grading system measures a) abnormality of the prostate cancer cells, b) histological growth pattern, and c) the speed of the spread of cancer. It remains the most powerful grading system since 1960. Xu et al. \cite{105} automated Gleason grading of prostate ADC using SVM algorithm and obtained AUC of 0.87. They extracted H channel from H\&E stained RGB images and applied multi-thresholding approach to remove the background. In \cite{76}, Lawson et al. used persistent homology to capture architectural features of Gleason patterns after performing Reinhard color normalization and color deconvolution operation. They used various statistical techniques including principal component analysis, hierarchical clustering and t‐distributed stochastic neighbor embedding to classify images according to the grading system. Linkon et al. \cite{107} shed light on the evolution of deep learning in prostate ADC detection and Gleason grading. Their extensive review of the literature highlighted lack of reproducible data in this research field. Several researchers automated the grading system \cite{74, 75, 104, 106} using DL models. Pantanowitz et al. \cite{74} reported ensemble CNN model for carcinoma diagnosis and grading; they performed tissue detection, classification, and slide level analysis. Gradient boosting classifier was used to differentiate tissue and background. AUC around 0.99 was obtained for external and internal test sets. Arvaniti et al. \cite{75} automated Gleason grading system using transfer learning approach on TMA dataset. The authors used different CNN models in which MobileNet performed well, with an average recall of 70\% for Gleason grades classification. Ambrosini et al. \cite{104} implemented CNN model to extract cribriform growth pattern of prostate ADC, achieving 0.81 AUC. Another research group \cite{106} used CNN model to detect ADC, grading them into different Gleason grades using 85 images collected at 20X magnification. They achieved accuracy of 91.05\% for benign versus ADC classification and 85.4\% for Gleason grading.

Gland segmentation helps detection of prostate cancer. Singh et al. \cite{77} applied pixel- and object-based approaches to segment different regions from tissue samples. They collected the images from TCGA database and digitized at 40X magnification. Rashid et al. \cite{78} differentiated benign and malignant glands based on nuclei feature and the ratio between epithelial layer area and lumen area of segmented regions. They used LDA for gland pixel labeling and watershed segmentation algorithm for ROI segmentation, followed by SVM for classification. A sensitivity of 0.83 and specificity of 0.81 were achieved for classification of glands.

\subsection{Female genital organs ADC and esophagus ADC}
\label{adc4}
Endometrial \cite{79} and ovarian \cite{80} tissue types were classified into benign and ADC classes. The classification was performed using CNN and attention mechanism, where VGGNet was the backbone \cite{79}. The network accuracy was around 76\% for four class classification and around 93\% for two-class classification using 498 tissue samples. Patch-based endometrium and colon ADC was detected by Riasatian et al. \cite{110}. They used the DenseNet topology as a feature extractor and performed classification using SVM. In \cite{80}, Yu et al. reported transfer learning approach on different CNN models, in which VGGNet identified the cancerous regions with AUC greater than 0.95.The authors also performed tumor grade classification and obtained AUC greater than 0.80. Tomita et al. \cite{81} extracted grid-based features using CNN and detected esophagus ADC through an attention-based technique. They used 180 WSIs, digitized at 20X magnification, and achieved mean accuracy of 0.83. The proposed model depends on tissue-level annotations, unlike contemporary methods that are based on pixel/ROI annotation \cite{10, 14, 22, 25, 26, 27, 33, 34, 35, 36, 40, 49, 52, 60, 68, 69, 70, 71, 72, 74, 75, 76, 77, 78, 82, 84, 86}. 

\subsection{Lung ADC}
\label{adc5}
As discussed in Section \ref{scc2}, lung carcinoma is categorized into four different subtypes. Several researchers proposed AI-based approaches to detect and classify lung cancer into its subtypes \cite{10, 11, 38, 39, 40, 41, 43, 44, 45, 46, 81, 82, 83, 84, 85, 86, 87, 88, 89, 94, 112}. Wang et al. \cite{82} segmented ADC nuclei using active contour method and classified ADC tumor cells, stromal cells, and lymphocytes using CNN model. The accuracy of 90.1\% was obtained for test dataset \cite{82}. Wen et al. \cite{83} performed nuclei segmentation using intensity and texture features. They used SVM and RF models to classify image patches as good quality, under quality, and over quality segmentation. RF achieved better accuracy than SVM for test datasets. F1 scores above 75\% were achieved for two independent datasets. Kanavati et al. \cite{84} differentiated carcinoma and benign tumor based on EfficientNet model in sliding window fashion. The authors obtained AUCs above 0.97 for four independent test datasets. Shi et al. \cite{85} proposed a pairwise-based deep ranking hashing algorithm to detect carcinoma image patches, achieving 97.4\% accuracy. Wang et al. \cite{86} performed an image-level classification of WSIs using a weakly supervised learning approach. A fully convolutional network was used to retrieve discriminative blocks and representative features from images. Further, they classified lung cancer as normal, SCC, ADC, and small cell carcinoma using RF classifier. Frank et al. \cite{112} performed image entropy tiles analysis using a customized CNN model. They obtained an accuracy of 92\% for lung ADC and SCC classification.

ADC was classified into its subtypes based on histological growth patterns using WSIs \cite{10, 87}. Nguyen et al. \cite{10} reported an accuracy of 89.24\% for classification of five ADC subtypes using CNN with soft voting technique. Wei et al. \cite{87} implemented ResNet model for carcinoma region identification. A total of 422 WSIs digitized at 20X magnification was considered for the study. A Kappa score of 0.525 and an agreement of 66.6\% with the three pathologists for classifying ADC subtypes was reported. Antonio et al. \cite{88} used pretrained autoencoder as a feature extractor for convolutional classifier and classified  transcriptome subtypes of lung ADC. They evaluated the results with three varied input images sizes and filter sizes. The sparse autoencoder network with 2048 × 2048 input size resulted in 98.9\% classification accuracy.

\subsection{Miscellaneous ADC}
\label{adc6}
Canine mammary tumors (CMTs) in dogs have elevated rates of occurrence and mortality, which include ADC as one of the subtypes. They are also considered great models for studies of human breast cancer. Kumar et al. \cite{89} evaluated a fused framework, where they extracted features using VGGNet and performed classification using SVM and RF on canine mammary tumor and human breast cancer datasets. The proposed framework with SVM resulted in mean accuracy of 97\% and 93\% for binary classification of human breast cancer and CMT, respectively. However, the highest accuracy was obtained at the lowest magnification. 

\section{Discussion and Future Directions}
\label{sec7}
Histopathological image analysis has become a recognized field because of the complexity involved in the analysis. Hence, there is a clear need for quantitative image-based assessment. This article presents a review on detection and classification of carcinomas using different approaches in CAD systems. Traditional ML follows five major steps, namely data collection, image preprocessing, segmentation of ROI, extraction of features from the segmented regions, and classification of tissue samples or detection of carcinoma. However, DL approaches learn hierarchical feature representations from the raw images. Some traditional image preprocessing operations such as image enhancement, color normalization, and noise removal from images suppress the unwanted data and enhance the image features. Various algorithms, namely Reinhard method, Macenko method, stain deconvolution, CLAHE, and Gaussian blur smoothing were applied before ROI segmentation \cite {75, 59, 26, 30, 61, 55, 58, 51, 65, 22, 19, 18, 21, 89, 23, 76, 15, 25, 62, 77, 52, 40, 41, 113, 105, 114, 109, 97, 96}. 

Some studies in the literature used segmentation methods to extract the ROI. Based on methods used in the literature, we categorized the segmentation methods into threshold-based, contour- and transform-based and cluster-based methods. In threshold-based approaches, use of Otsu\textquotesingle s thresholding \cite{75,26, 30, 71, 65, 21, 76, 77, 40, 38, 41, 95, 92, 114, 109}, incremental thresholding \cite{51, 52}, and multiple thresholding \cite{20, 105} were reported for extraction of ROI. Contour-based segmentation includes active contour methods \cite{28, 51, 82} and Sobel edge detectors \cite{21}. Transform-based approach mainly used watershed segmentation \cite{18, 23, 29, 72, 70, 86}. Cluster-based segmentation includes k-means clustering, fuzzy c-means clustering, and DBSCAN \cite{69, 32, 56, 15, 97}. The literature also includes single-pass voting \cite{11}, region-based segmentation \cite{60,61}, graph-based segmentation \cite{59}, deep network segmentation models \cite{98, 102}, and connected component analysis \cite{47}. However, majority of the research groups performed segmentation using thresholding. Various types of texture features, topological features, morphological features such as shape, structure, color, pattern and size, Z-transform features, object-based features, and CNN features were extracted from ROI or image patches. Feature extraction was performed either on grayscale images or on specific color components of RGB, LAB, and HSV/HSI \cite{54, 30, 71, 20, 33, 61, 55, 36, 51, 21, 23, 69, 76, 32, 39, 29, 56, 15, 72, 49, 63, 62, 25, 78, 70, 43, 77, 52, 82, 83, 66, 67, 38, 41, 53, 114, 102}. We summarized the extracted features reported in the literature in Tables \ref{tab2} and \ref{tab3}.

\begin{figure*}
    \centering
    \includegraphics[width=95mm,scale=1.9]{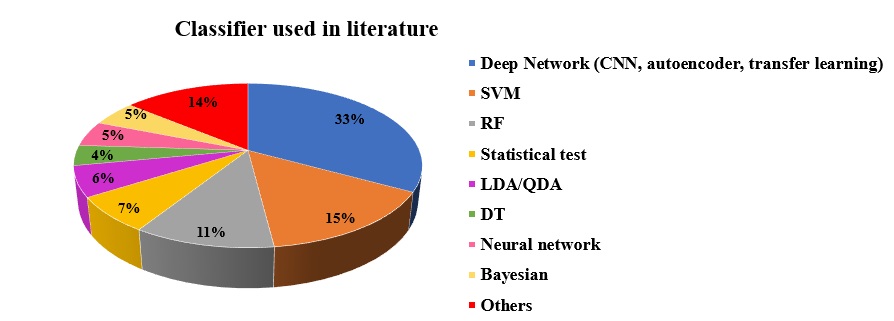}
    \caption{Classifiers used in literature}
    \label{figure8}
\end{figure*}

The state-of-the-art literature considered various classifiers for carcinoma detection and classification, as listed in Tables \ref{tab2} and \ref{tab3}. Fig.\ref{figure8} shows the percentage of various classifiers used by researchers in the literature. We highlighted the top three classification techniques in which deep neural networks \cite{75, 59, 46, 26, 28, 33, 10, 14, 64, 58, 84, 44, 65, 74, 27, 68, 85, 79, 81,  40, 82, 87, 50, 67, 80, 92, 94, 95, 96, 98, 100, 102, 104, 106, 107, 108, 109, 110, 112, 113, 114, 115} such as pretrained models (AlexNet, VGGNet, ResNet, InceptionNet, EfficientNet, etc.), autoencoders, and customized CNNs top the list. SVM \cite{89, 56, 72, 78, 77, 93, 96, 105, 110} and RF tree classifier \cite{89, 69, 72, 70, 68, 86, 93, 114} were second and third in the list. In the `Others' category, we included hashing, LR, quasi-supervised learning algorithm, ensemble learning, and linear classifier based on different features extracted from ROI for decision-making.

\begin{figure*}
    \centering
    \includegraphics[width=115mm,scale=1.9]{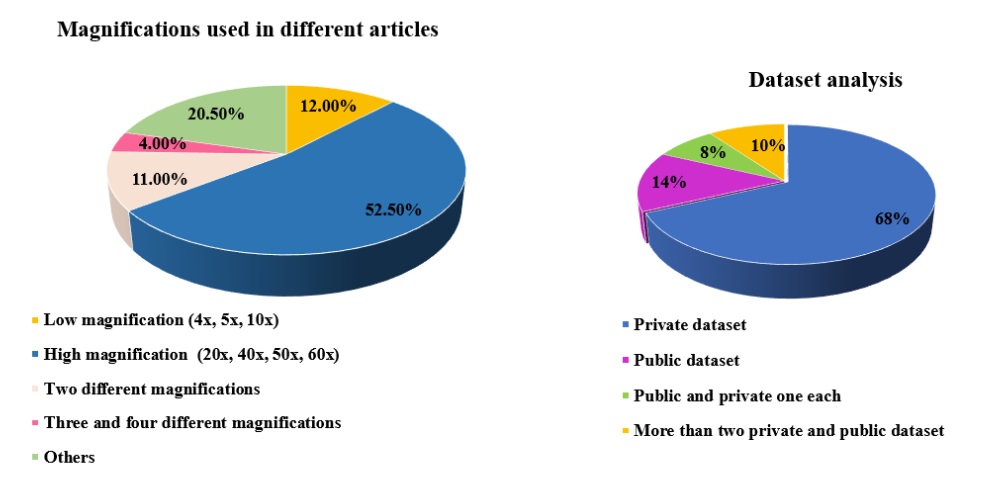}
    \caption{Different image magnifications used in the literature (left) Dataset used in the articles (right)}
    \label{figure9}
\end{figure*}

We listed different magnifications used to capture the traditional microscopic images and digitized WSIs in Tables \ref{tab4} and \ref{tab5}. Additionally, some studies reported  hierarchical cascaded scheme for detecting ADC on digitized histopathology images at different scales \cite{65, 71}. Fig.\ref{figure9} (left) shows that 52.5\% of the articles detected and classified carcinoma using high magnification images. It indicates that more than 50\% of researchers used 20X, 40X, 50X, and 60X magnification images. We studied the usage of private and public datasets reported in the literature. Out of 101 articles, 68\% of the articles used private datasets, and around 8\% used both public and private datasets as shown in Fig.\ref{figure9} (right).

Performance measures are used to evaluate ML and DL models. When we observe the literature, there were various evaluation metrics reported for ADC and SCC classification. In Tables \ref{tab6} and \ref{tab7}, we listed evaluation measures used in SCC and ADC classification, respectively. Among various evaluation metrics, accuracy was the commonly used performance measure. In the case of some studies, the lower performance could be due to small datasets with high magnification images \cite {22, 54, 78, 61, 68, 93}, imbalanced dataset \cite{79, 81}, CNN with small datasets \cite{28}, down-sampled images, and WSI scanning artifacts \cite{14}. In earlier days, only cells and nuclei shape features were used for classification \cite{20}. However, statistical features, Haralick texture features, fractal dimension, and color features were used in later days \cite{15, 22, 25}. A few researchers used small dataset and achieved good results \cite{25, 37, 30, 33, 47, 49}. This may be because the datasets were not representative of different microscopic features. It could also be because images were captured at low magnification but covering large areas of views \cite{16, 19}. Some researchers applied augmentation techniques to increase the dataset size \cite{26, 27, 28, 50}, which improved the model accuracy. Further, most of the studies of lung cancer subtype classification were either performed on TCGA, Stanford TMA and ICGC WSI datasets or considered both private and public WSIs to train the model \cite{46, 40, 38, 11, 44, 43, 45, 94, 95, 100}. However, the qualities of histopathology slides from TCGA and hospitals were quite different. Hence, we could observe a variation in model performance \cite{114}. In a few studies \cite{93, 96}, the model performance was less as compared to other lung cancer classification models because of microscopic images from private hospitals with small sized datasets. The studies \cite{11, 38, 41} reported the use of only cell-level information from the images captured at 40X magnification for classification, which resulted in model accuracy of around 85\%. Weak boundaries of the cell images and strong edges within the cells can mislead the classification process. When we use the hashing method for classification with limited number of hash bits,it may result in same hash entries for different images \cite{11}. In the case of small dataset, traditional ML approach achieved better results than CNN model \cite{68}, and combining two datasets resulted in better performance \cite{113}.

\begin{table*}[hbt]
\setlength{\tabcolsep}{1.5pc}
\caption{Performances metrics measured in SCC article}
\label{tab6}
\begin{tabular*}{\textwidth}{@{}ll}
\hline Ref, Year & Performance metrics \\
\hline
 \multicolumn{2}{c}{Head and Neck} \\
 \hline
\cite{20}, 2003 & Acc: Normal vs. dysplasia 63.7\% Normal vs. SCC  82.7\% Dysplasia vs. SCC 73.3\%  \\
\cite{37}, 2007 &	Acc: 96\% \\
\cite{19}, 2012 & Acc: 95.70\%, Sen: 94.70\%, Spec: 98.80\% \\
\cite{22}, 2012 & Acc: 91.64\%, Sen: 94.17\%, Spec: 91.40\% \\
\cite{16}, 2015 & Sen: 0.98, Spec: 0.99, PPV 90.00\% \\
\cite{30}, 2017	& Acc: 96.92\%, Sen: 95\%, Spec: 97.35\%, AUC: 96.45\%, F-score: 92\%, Prec: 89\% \\
\cite{25}, 2018 & Acc: 89.7\%  and 100\% \\
\cite{14}, 2019 &	Acc: 84.8 ± 1.6\%, Sen: 84.7 ± 2.2\%, Spec: 85.0 ± 2.2\% , AUC: 0.916 -0.944, F-score: 84.8 ± 1.5\% \\  
\cite{27}, 2019	& Acc: 96.31\%, Sen: 92.66\%, Spec: 97.55\%, AUC: 98.91\%, F-score: 92.62\% \\
\cite{28}, 2019 & Sen: 88.87\%, Prec 82.03\% \\
\cite{33}, 2019 & Acc: 97.05\%, Sen: 84.74\%, Spec: 98.64\%, JI: 77.21\%, Dice: 87.14\% Prec: 89.35\% \\
\cite{15}, 2020 & Acc: 95.56\%, Sen: 0.91, Spec: 0.97, F-score: 0.91, Mathew’s coef: 0.088  \\
\cite{26}, 2020 &	Acc: 97.51\%, Sen: 93.67\%, Prec: 97.14\% \\
\cite{98}, 2021 &	AUC: 0.96 \\
\hline
\multicolumn{2}{c}{Lung} \\
\hline
\cite{41}, 2002 & Err 11.6\%, Err (false negative) 2.7\%, Err (false positive) 4.5\% \\
\cite{40}, 2013 &	Dataset1 Acc: 92.41\%, Dataset2  Acc: 95.42\% \\
\cite{11}, 2015 &	Acc.: 87.30\% \\
\cite{38}, 2016 &	AUC: 0.85 \\
\cite{43}, 2017 &	Acc: 97.98\%, Prec: 97.50\% \\
\cite{44}, 2018 &	Dataset1 Acc: 92\%, Dataset2 Acc: 91\%  \\
\cite{46}, 2018 &	AUC: 0.97 \\
\cite{45}, 2020 &	AUC: 0.883-0.932 \\
\cite{94}, 2020 &	Cancer types: AUC 0.995 ± 0.008, Subtype: AUC 0.87 ± 0.1 \\
\cite{96}, 2021 &	Acc :ADC Vs. SCC 73.91\% , SCC vs small cell carcinoma 83.91\% \\
\cite{92}, 2021 &	AUC: 0.94 to 0.99 \\
\cite{95}, 2021 &	Four medical centres: Acc: 0.860, 0 .870, 0.890, 0.8000, AUC: 0.970, 0.918, 0.963, 0.978  \\
\cite{100}, 2021 &	AUC: 0.9414 - 0.9594 \\
\hline
\multicolumn{2}{c}{Cervix} \\
\hline
\cite{54} 2017 & Acc: 77.25\% \\
\cite{50} 2018 & Acc: Before aug. 89.48\% after aug. 93.33\% \\
\cite{99} 2021 & Acc: 90.00\%, Sen 0.94, Spec 0.97, F-score: 0.92, Precision 0.91 \\
\hline
\multicolumn{2}{c}{Skin} \\
\hline
\cite{47}, 2016 & Acc.: 83.3\%, Sen:  84.6\%, Spec:  81.8\%, Prec: 84.6\% \\
\cite{49}, 2016 &	Acc: 85.18\%, Sen: 91.66\%, Spec : 80.00\%, Prec: 78.57\% \\
\hline
\multicolumn{2}{c}{Esophagus} \\
\hline
\cite{55}, 2019 & 97.78\% ± 0.05, Sen 100±0, Spec. 95.55±10, Mathew’s coef 0.96±0.09, Kappa score 0.96±0.10 \\
\hline
\end{tabular*}
\end{table*}
 
\begin{table*}[hbt]
\setlength{\tabcolsep}{1.5pc}
\caption{Performances metrics measured in ADC article}
\label{tab7}
\begin{tabular*}{\textwidth}{@{}ll}
\hline Ref, Year & Performance metrics \\
\hline
 \multicolumn{2}{c}{Colon} \\
 \hline
\cite{61}, 2010 & Acc: 87.59 ± 5.01, Sen: 85.80 ± 6.71, Spec: 89.14 ± 10.40, Dice: 88.91 ± 4.63 \\
\cite{58}, 2017 & Acc: Above 95\%, Sen: 0.93, F-score: Above 0.96, Prec: 0.964 \\
\cite{115}, 2020 & Acc: 96\%, Sen: 0.945, Spec: 0.975, F: score: 0.9594, Prec: 0.974 \\
\cite{64}, 2020	& AUC: 0.96 and 0.99 \\
\cite{113}, 2021 & Two datasets and combined Acc: 96.16\%, 92.83\% and 99.13\%, Sen: 92.83\%, 93.11\% and 97.85\% \\
\cite{114}, 2021 & Acc: Dataset1 94.6\%, Dataset2 92.0\% \\
\hline
\multicolumn{2}{c}{Gastric and pancreas} \\
\hline
\cite{68}, 2017 & Acc: 0.6990 to 0.74 \\
\cite{65}, 2020 & Acc: 0.937, Sen: 0.939, F-score: 0.934, Prec: 0.929 \\
\cite{64}, 2020 & AUC: 0.97 and 0.99 \\
\cite{109}, 2021 & Sen: 0.9705, Spec: 0.9272, AUC: 0.9166, Dice: 0:8331 \\
\cite{108}, 2021 & Acc: 0.9417, Sen: 0.9302, Spec: 0.9706, AUC: 0.984 \\
\hline
\multicolumn{2}{c}{Prostate} \\
\hline
\cite{71}, 2006 & Acc: 88\% \\
\cite{78}, 2013 & Sen: 0.83, Spec: 0.81 \\
\cite{73}, 2019 & Sen: 87.1\%, Spec: 90.7\%, AUC: 0.951 \\
\cite{70}, 2019 & Acc: 93\%, Sen: 90\%, Spec: 93\% \\
\hline
\multicolumn{2}{c}{Lung} \\
\hline
\cite{85}, 2018 & Acc: 97.49\%, Prec 97.49\% \\
\cite{86}, 2019 & Two datasets: Acc: 97.3\% and 82.0\% \\
\cite{84}, 2020	& AUC: Four independent test set: 0.975, 0.974, 0.988 and 0.981 \\
\cite{112}, 2021 & Acc: 92\% \\
\cite{93}, 2021 & Acc: 78.33\% \\
\hline
\multicolumn{2}{c}{Overy, Endometry and Esophagus} \\
\hline
\cite{79}, 2019 & Acc: 84.50\%, Sen: 77.97\%, Spec: 100\%, AUC: 0.9829 \\
\cite{81}, 2019 	& Acc: 0.83, Sen: 0.60, F-score: 0.59, Prec: 0.62 \\
\cite{80}, 2020 & AUC > 0.81 \\
\hline
\end{tabular*}
\end{table*}

\begin{table*}[hbt]
\caption{Summary of SCC articles}
\label{tab2}
\begin{tabular}{p{1.5cm}p{3.5cm}p{3cm}p{3.5cm}p{3.5cm}}
\hline SCC organ name & Preprocessing	& Segmentation & Features Extracted & Classification \\
\hline
Head and neck region  & CLAHE \cite{15}, Low pass filter \cite{18, 19}, Contrast enhancement \cite{18, 19, 25, 97}, Histogram specification and equalization \cite{21, 22}, Color deconvolution \cite{23}, Gaussian smoothing \cite{26}, Contourlet transform \cite{28}, SVD \cite{30} & k-means \cite {15, 97}, Watershed \cite{18, 19, 23, 29}, Thresholding \cite{20, 21, 26, 30}, Sobel edge detector \cite{21}, Neural network \cite{22}, Chan-Vese \cite{28}, DBSCAN \cite{32}, UNet and SegNet\cite{102} & Various texture feature \cite {15, 18, 19, 21, 22, 25, 27, 29, 30, 32, 97}, Morphological features \cite {15, 18, 32, 20, 29, 35, 36, 97}, Epithelial measurement \cite{21, 23}, Fractal dimension \cite{20, 22, 30}, Hu\textquotesingle s moments \cite{30} & Transfer learning models \cite{26, 27, 14, 98}, Custom CNN \cite{26, 28, 33}, Machine learning algorithms \cite {15, 19, 20, 22, 23, 25, 29, 30, 32}, Statistical test \cite {21, 34, 35, 36, 37}\\ 
Lung &	cw-HE \cite{40}, Gaussian smoothing \cite{41, 96}, Adaptive histogram equalization \cite{96} & Single pass voting \cite{11}, Thresholding \cite{38, 40, 41, 92, 95} & Various texture feature \cite{38, 39, 40, 93, 11, 43, 96}, morphological features \cite{38, 39, 40, 41}, Densitometric features \cite{40}, ConvNet features \cite{43, 96}, Homology-based features \cite{93}, Wavelet features \cite{96} & Hashing-based majority voting \cite{11}, Machine learning algorithms \cite{38, 93, 96, 100}, Statistical test \cite{39}, Neural network \cite{41, 44, 93, 96}, Ensemble classifier \cite{40}, Graph-based model \cite{43}, Transfer learning models \cite{44, 45, 46, 94, 92, 95, 100} \\
Cervical & Smoothing filter, followed by histogram equalization \cite{51}, CLAHE \cite{52} & Thresholding \cite{51, 52}, Graph search  \cite{53} & Morphological features \cite{51, 53}, Statistical, structural and spectral features \cite{52}, ConvNet features \cite{54, 99} & Transfer learning model \cite{50}, Statistical test \cite{51}, Machine learning algorithms \cite{52, 54}, Ensemble network \cite{99} \\
Skin &	- & Connected components analysis \cite{47}	& Intensity features \cite{47}, Z-transform features \cite{49} & Machine learning algorithms \cite{47, 49} \\
Esophagus &	CLAHE \cite{55}	& - & Histogram variations, epithelial thickness, Morphology features \cite{55} &	Machine learning algorithms \cite{55} \\         
\hline
\end{tabular}
\end{table*}

\begin{table*}[hbt]
\caption{Summary of ADC articles}
\label{tab3}
\begin{tabular}{p{1.5cm}p{3.5cm}p{3cm}p{3.5cm}p{3.5cm}}
\hline
ADC organ name & Preprocessing	& Segmentation & Features Extracted &	Classification \\
\hline
Colon, Stomach and Pancreas & CLAHE \cite{58}, sparse stain separation \cite{59}, Stain deconvolution \cite{62, 109}, Gaussian blur smoothing \cite{65, 114} & k-means \cite{56, 61}, Graph-based \cite{59}, Dense UNet \cite{59}, Region growing \cite{60, 61}, Thresholding \cite{65, 114, 109}, CNN \cite{67} & Zoom-out features \cite {56}, ConvNet features \cite{56, 62, 110, 114}, Object graph features \cite{61}, Topological features \cite{62}, Various texture features \cite{63, 67, 68}, Morphology features  \cite{66, 67, 68}	& Transfer learning models \cite{63, 68, 108}, Machine learning algorithms \cite{56, 61, 62, 68, 114, 110}, Custom CNN \cite{58, 65, 68, 115}, Quasi-supervised learning \cite{63}, Neural network \cite{64, 65, 109}, Ensemble CNN \cite{113} \\
Prostate &	Color Deconvolution \cite{73, 105}, Gaussian filtering \cite{75, 77}, Reinhard color \cite{76} &	Watershed \cite{70, 72, 78}, Thresholding \cite{71, 75, 77, 105} &	Morphological features \cite{70, 72, 73, 77}, Various texture features \cite{71}, Wavelet features \cite{71}, SURF descriptors \cite{72}, Topological features \cite{76}, Epithelial layer architecture features \cite{78}	& Machine learning algorithms \cite{70, 72, 77, 78, 105}, Ensemble classifier \cite{71, 73}, Ensemble CNN \cite{73}, Transfer learning models \cite{75, 106}, Hierarchical Ward clustering \cite{76}, Custom CNN \cite{104} \\ 
Lung & Reinhard method \cite{10} & Active contour \cite{82}, Thresholding \cite{86} &	Morphological features \cite{84}, ConvNet features \cite{88} & Transfer learning models \cite{12, 89, 112}, Custom CNN \cite{84, 87, 112}, RF \cite{86}, Autoencoder \cite{89} \\
Ovary, Endometrial \& Esophagus &	- & - &  ConvNet features \cite{110}	& Transfer learning models \cite{81, 82, 83}, Custom CNN \cite{81}, SVM \cite{110}\\
ADC in dog and mice & Macenko method \cite{89} & k-means \cite{69}, Watershed \cite{69} & Object level features \cite{69}, ConvNet features \cite{89} & Machine learning algorithms \cite{69, 89} \\
\hline
\end{tabular}
\end{table*}

\begin{table*}[hbt]
\caption{Number of microscopic images with magnification details reported in the articles}
\label{tab4}
\begin{tabular}{p{3cm} p{3cm} p{2cm} p{2cm} p{2cm} p{3cm}}
\hline
\multirow{2} {*}{Magnifications} &
\multicolumn{5}{m{12cm}}{Number of images in range (Train + validation + test)}  \\ 
& 0-200	& 200-400 & 400-600 & 600-800 & Image patches \\
\hline
Low magnification (2.5X, 4X, 5X, 10X) &	\cite{18}, \cite{19}, \cite{21}, \cite{93} &\cite{20}, \cite {63}, \cite{98}	& - & - & \cite{93} (25K)\\
High magnification (20X, 40X, 45X, 50X, 60X) & \cite{53}, \cite{61}, \cite{106}, \cite{22}, \cite{49}, \cite{47}, \cite{77} &	\cite{28}, \cite{25} &	\cite{23}, \cite{108}, \cite{99}, \cite{50} &	\cite{15}, \cite{97} &	\cite{113} (1lakh), \cite{85} (5256), \cite{41} 552 cells image \\
Both low and high magnifications & \cite{55} & \cite{89} & - & - & \cite{46} (1634), \cite{82} (1337) \\
No magnification &	\cite{96} & - & - &	-&\cite{115} (9200) \\
\hline
\end{tabular}
\end{table*}

\begin{table*}[hbt]
\caption{Number of WSIs and TMA with digitized magnification details reported in the articles}
\label{tab5}
\begin{tabular}{p{3cm} p{3cm} p{2cm} p{2cm} p{2cm} p{3cm}}
\hline
\multirow{2}{*}{Magnifications} &
\multicolumn{5}{m{12cm}}{Number of images in range (Train + validation + test)}  \\ 
& 0-200	& 200-400 & 400-600 & 600-800 & Between 1000-4000 \\ \hline
Low magnification (2.5X, 4X, 5X, 10X) & - &	\cite{105}	& - & - & \cite{86}, \cite{92} \\
High magnification (20X, 40X, 45X, 50X, 60X) & \cite{27}, \cite{32}, \cite{33}, \cite{56}, \cite{59}, \cite{76}, \cite{73}, \cite{78}, \cite{81}, \cite{26}, \cite{30}, \cite{37}, \cite{58}, \cite{62}, \cite{68}, \cite{71}
\cite{69} &	\cite{10}, \cite{14}, \cite{40} &	\cite{87} & \cite{75} &	\cite{64}, \cite{84}, \cite{100}, \cite{38}, \cite{44}, \cite{114}, \cite{109}, \cite{95}, \cite{74} \\
Both low and high magnifications & \cite{70}, \cite{52},
\cite{65} &	\cite{102}  & \cite{79}, \cite{72} & -&	\cite{66}, \cite{39}, \cite{67} \\
No magnification &  \cite{104}, \cite{112} & -& - & - & -\\
Number of cases & \cite{34}, \cite{36}, \cite{35}, \cite{60}, \cite{115}& -& - & - & -\\
\hline
\end{tabular}
\end{table*}

In Tables \ref{tab2}, \ref{tab3}, \ref{tab4}, and \ref{tab5} we summarize magnification details, various traditional image processing steps, and ML and DL techniques used to automate SCC and ADC diagnosis from histopathological images. In this section, we discuss the overall outcome of our review of literature and future direction in AI-based techniques for carcinoma diagnosis that needs consistent collaboration of computer scientists, clinicians, and pathologists. 
\begin{itemize}
\item \textbf{Image preprocessing: }
Some of the issues that affected SCC/ADC detection systems include variations in illumination, color shade variations, and image blurring. This is due to uneven staining,  tissue folds, or debris. Such artifacts can be handled by using different preprocessing algorithms. Kumar et al. \cite{89} performed a study to understand the color normalization algorithms on ConvNet feature extractor. Their results showed that the accuracy without stain normalization was better than that with stain normalization. However, Ozturk et al. \cite{91} classified the image as cancerous and noncancerous patches after applying noise reduction and enhancement techniques for deep networks. The authors reported that the preprocessing methods contributed to the learning to some extent. However, excessive preprocessing decreased the model performance. From the above findings, we can observe that the researchers$'$ opinions are divided on the need of preprocessing for DL models. However, it is a required step in traditional ML methods to extract appropriate handcrafted features. To illustrate, the authors reported that the ML-based skin SCC detection system\textquotesingle s accuracy was reduced because of not using any preprocessing techniques \cite{49, 47}.  Presently, researchers use the color augmentation technique as one of the data augmentation techniques to increase the training dataset. Nevertheless, excessive color changes can lead to misclassification. 

\item \textbf{Generalizability: }
Several researchers considered tissue samples from specific organs of origin of SCC and ADC. They implemented various ML algorithms and CNN models to detect carcinomas. However, the state-of-the-art methods lack generalizability. To date, even though many automated systems have been developed, they are not suitable for being applied to a similar problem. For example, whether a diagnostic model developed for SCC/ADC of lung is useful for diagnosing SCC/ADC of esophagus has not yet been explored. Hence, we need a generalized model that eliminates need for multiple decision systems for carcinoma diagnosis based on primary organs. The robustness of the automated system could be improved by including a variety of data during training.


\item \textbf{Effect of magnification: }
 Generally, a pathologist scans tissue samples under a microscope at multiple magnifications for interpretation and diagnosis. Low magnification images include a larger field of view as compared to high magnification images. We observe that the researchers used different magnifications for their studies to extract the features. For instance, Hosseini et al. \cite{55} trained a classification model separately by taking the images from 10X and 20X magnifications and proposed a fractal-based approach to differentiate normal and esophagus dysplasia. Kumar et al. \cite{89} studied the influence of magnification in carcinoma classification. They achieved higher accuracy at 4X and 10X magnification and lower accuracy at 40X magnification. Further, there are a few studies listed in Tables \ref{tab4} and \ref{tab5} that considered images of two or three different magnifications to train the model. However, no studies reported the feature extraction from images captured at multiple magnifications. Microscopic images from multiple magnifications could be used to improve classification and detection results. Additinally, images extracted from WSIs at multiple magnifications may be used for this purpose. Extracting different microscopic features from images at multiple magnifications would more closely mimic the pathologists$'$ evaluation.

\item \textbf{Ground truth: }
It is evident from the review of literature that the majority of algorithms proposed were based on fully supervised learning. Recently, a few research groups proposed weakly supervised learning techniques \cite{84, 86} for fast classification with a small number of ROI annotated images and the rest all were image-level labels. Obtaining well annotated data for supervised learning often requires domain expertise. For instance, in \cite{84}, a group of 39 pathologists annotated the WSIs. There may be inter-observer variability that should be quantified among the pathologists. In \cite{75, 72, 87}, inter-observer variability was measured using the Kappa coefficient. The major problem in supervised learning is to obtain the ground truth for training a model. Most of the articles considered private datasets and some used public datasets, namely TCGA and TCIA of WSIs. However, none of the public datasets provided ROI annotation with original histopathological images for SCC or ADC detection. There is a lack of standard datasets with annotation to validate and compare the proposed algorithms.

\item \textbf{Pretrained networks and deep networks: }
It is evident in this article that out of 101 research papers, vast majority of contributions by deep networks were published after 2017. Some of the studies used pretrained networks as feature extractors \cite{43, 54, 56, 62, 88, 89, 96, 99, 110, 114} and classified carcinoma into different subtypes. These pretrained networks were trained on natural scene images (e.g., ImageNet). The features learned from these networks may not be accurate for histopathological images. Given a small dataset, training only the last few decision layers using the regularization technique is recommended to avoid overfitting. In the last two years, end-to-end trained networks for carcinoma diagnosis were proposed by many researchers \cite{26, 28, 33, 58, 65, 68, 81, 84, 87, 104, 112, 115}. However, there is no clear way to select and design CNN models from scratch. Hyperparameter optimization (decay, dropout rate, learning rate, etc.) squeezes additional performance from a deep network model. It would be more useful to have a clear recipe to obtain the best set of hyperparameters. Many researchers chose hyperparameters randomly, which often seems to work well. However, some tips for tuning hyperparameters for histopathological images would help researchers to obtain more accurate and faster results. A few disadvantages of DL are 1) training deep networks from scratch that requires a large amount of data to obtain superior results, 2) model requires huge memory and is computationally expensive, and 3) accountability is important in medical image analysis otherwise it becomes a serious legal issue. Hence, it is necessary to observe the intermediate layers result of deep networks in order to understand the complex feature patterns.

\item \textbf{Whole slide image: }
Carcinoma diagnosis using WSI is a trending topic. During our review process, we could find more than 50\% of articles that used WSIs for detection and classification ( details \ref{tab4} and \ref{tab5}). A pathologist usually takes around fifteen minutes to evaluate one whole slide \cite{86}. Successful application of DL models in WSI analysis was described in the literature. However, a major issue faced during WSI analysis is that it is hard to train DL models directly because of its huge file size and GPU/CPU memory requirement. Instead, most of the reported studies extracted small patches with the help of pathologists, divided the images into random patches, quantized images from higher bit to lower bit representation, or resized the images. These steps may lead to information loss from the images or may take a long time to extract the patches. A reasonable strategy would achieve minimal information loss by utilizing maximal architecture capacity and minimum human intervention. Several research groups used varied patch sizes ranging from 32x32 pixels to 10,000x10,000 pixels. The patch selection itself is a key research area for WSI analysis. Currently, multiple scanners from different vendors output the images in different file formats with varied pixel resolutions. The WSI size is depends on the scanner\textquotesingle s objective lens magnification and number of individual pixels within the digital camera\textquotesingle s sensor and the monitor. Hence, comparison among the results of the literature across the various studies is a challenging task, which could be an interesting future research direction.

\item \textbf{Dataset: }
The quality of trained model depends on the quantity of data in the dataset. Imbalanced classes in the dataset may lead to biased results. In Tables \ref{tab4} and \ref{tab5}, we can observe the numbers of microscopic images and WSIs reported in the literature. Training CNNs using a small dataset may lead to overfitting. Image augmentation reduces the time and efforts  required to obtain additional training data, increases the robustness and generalization potential of CNN, and alleviates overfitting. Some techniques such as rotation, flipping, translation, transformation, scaling, and color jittering can be applied. Generative adversarial networks offer synthetic data augmentation. However, the techniques applied should not alter any information hidden in images. Researchers used different sizes of images to train their models. So far, there is no publicly available dataset that could be used to validate and compare algorithms$'$ performance. It is critical to know how big such datasets would have to be and what kinds of variations need to be covered in it. Some of the researchers used only publicly available databases, which could be different in terms of what pathologists encounter in their day-to-day practice. Hence, it is desirable to train the model using both private and public datasets.

\item \textbf{WSI versus microscopic images: }
The introduction of WSI helps to convert entire tissue on glass slides into a digital format that pathologists can view on a computer screen at different magnifications. They can also navigate spatially in the same way as a microscopic study using different computer software (e.g., Image Scope). The resolution of images differs greatly depending on the digital cameras used and the numerical aperture of the objective lenses used in the WSI systems. Fig.\ref{figure10} shows a few images at 20X magnification with different cameras. It is important to note that WSI varies from conventional microscopy image because of the camera\textquotesingle s sensor. We observe that many researchers proposed automated SCC/ADC detection systems using WSIs, whereas many other researchers used traditional microscopic images collected from clinical environments. Additionally, whole slide imaging involves a huge initial cost of the scanners, which is difficult to bear for small pathology laboratories. Hence, it is desirable to develop a robust approach that can handle both types of images.

\begin{figure}
    \centering
    \includegraphics[width=80mm,scale=1.0]{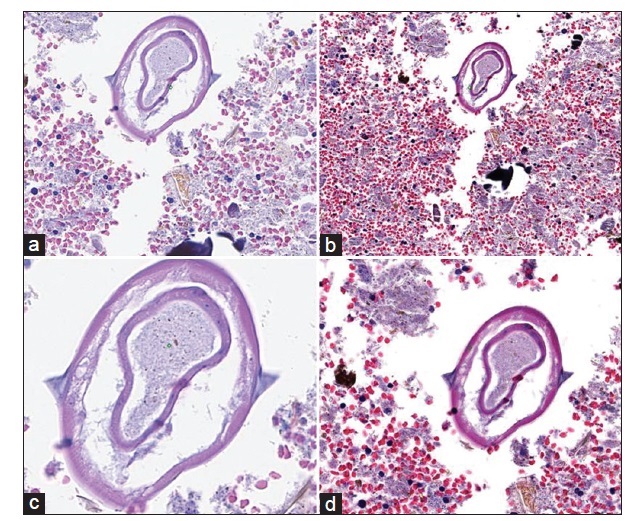}
    \caption{Images of an enterobius parasite using a similar 20X objective, but with different cameras \cite{90}}
    \label{figure10}
\end{figure}

\item \textbf{ML and DL approaches: }
Several studies investigated the potential of ML and DL that could detect and classify carcinomas. A comparative study of carcinoma detection and classification between deep neural networks and ML models is desirable. The results provided by ML-based approaches are verifiable, whereas deep neural models are certainly complex. Thus, DL and ML approaches could be integrated to produce a hybrid approach that provides a reliable and accountable solution. It is also interesting to combine evolutionary optimization with them \cite{nakane2020application}. 
\end{itemize}

\section{Conclusion}
\label{sec8}
This review highlights the massive opportunity for automated carcinoma (SCC/ADC) detection and classification. We organized the articles based on specific organs of carcinoma origin such as lung, head and neck region, female genital system, skin, prostate, colon, stomach, and esophagus. There is limited evidence of generalized carcinoma diagnostic system. Hence, it is necessary to develop novel methods to aid SCC/ADC diagnosis using modern AI-based approaches. Integration of different microscopic features obtained from images at multiple magnifications requires a comprehensive evaluation of each feature on a large multi-centric dataset. Most recent studies used deep neural networks and WSIs in carcinoma diagnosis. However, having a good prediction system is often not enough. Future avenues could include the development of explainable AI approaches which will aid early detection of SCC/ADC. Ultimately, automation brings down lead time in carcinoma diagnosis and also reduces pathologists$'$ workload. This would benefit hospitals that lack experts in pathology, giving them some initial insights into the diseases.





\bibliographystyle{elsarticle-num}

\bibliography{Review_Article.bib}  

\bio{}
\endbio


\end{document}